\documentclass[twocolumn, amsmath,amssymb,aps,prx]{revtex4-1}
\usepackage{verbatim}
\usepackage{graphicx}
\usepackage{ifpdf} % This lets us get figures right both in pdf and ps versions
\usepackage{dcolumn}  % Align table columns on decimal point
\usepackage{epsfig}
\usepackage{color}
\usepackage[usenames,dvipsnames]{xcolor}
\usepackage{siunitx}
\usepackage{soul} % to highlight text

\newcommand{\kT}{k_\mathrm{B}T}

\newcommand{\celsius}{\ensuremath{^\circ}C}

\def\eq{Eq.}

\begin{document}

\title{Forces between Silica Particles in Isopropanol Solutions of 1:1 Electrolytes}

% authors
\author{Biljana Stojimirovi\'c}
\affiliation{Department of Inorganic and Analytical Chemistry, University of Geneva, Sciences II, 30 Quai Ernest-Ansermet, 1205 Geneva, Switzerland}

\author{Marco Galli}
\affiliation{Department of Inorganic and Analytical Chemistry, University of Geneva, Sciences II, 30 Quai Ernest-Ansermet, 1205 Geneva, Switzerland}

\author{Gregor Trefalt}
\email{E-mail: \texttt{gregor.trefalt@unige.ch}}
\affiliation{Department of Inorganic and Analytical Chemistry, University of Geneva,
Sciences II, 30 Quai Ernest-Ansermet, 1205 Geneva, Switzerland}

\date{\today}

\begin{abstract}
Interactions between silica surfaces across isopropanol solutions are measured with colloidal probe technique based on atomic force microscope. In particular, the influence of 1:1 electrolytes on the interactions between silica particles is investigated. A plethora of different forces are found in these systems. Namely, van der Waals, double-layer, attractive non-DLVO, repulsive solvation, and damped oscillatory interactions are observed. The measured decay length of the double-layer repulsion is substantially larger than Debye lengths calculated from nominal salt concentrations. These deviations are caused by pronounced ion pairing in alcohol solutions. At separation below 10~nm, additional attractive and repulsive non-DLVO forces are observed. The former are possibly caused by charge heterogeneities induced by strong ion adsorption, whereas the latter originate from structuring of isopropanol molecules close to the surface. Finally, at increased concentrations the transition from monotonic to damped oscillatory interactions is uncovered.
\end{abstract}

\maketitle

\section{Introduction}

Forces between surfaces immersed in liquids are important in many natural and technological processes. We can find examples of such processes in biological systems, waste water treatment, ceramic processing, ink-jet printing, and particle design~\cite{Valle-Delgado2004,Govrin2017,Bolto2007,Cerbelaud2010,Kuscer2012,Zanini2017a}. Recent advancement in the force probing techniques such as surface force apparatus (SFA), colloidal probe technique based of atomic force microscopy (AFM), and optical tweezers enable routine surface force measurements with high precision and excellent reproducibility~\cite{Israelachvili2011,Trefalt2017b,Smith2020}.

A vast majority of the surface force measurements are done in aqueous systems. Some examples of such measurements aimed to study the effects of multivalent ions on electrostatic interactions~\cite{Zohar2006,Besteman2004,Moazzami-Gudarzi2016c} or mechanisms behind oscillatory structural forces~\cite{Zeng2011,Klapp2007,Moazzami-Gudarzi2016a}. Although water is the most important natural solvent, processes in non-aqueous media are equally interesting in view of technological as well as some natural processes. An example of such a process includes ceramics processing, where organic polar media, such as alcohols or ketones, are used for milling and homogenization of ceramic powder mixtures, which permit production of high-quality complex ferroelectric or structural materials~\cite{Rojac2014, Wang1997}. Another example of a process using non-aqueous solvents is printing of materials in 2D or 3D shapes. Material inks can be either completely non-aqueous based or can contain large portions of non-aqueous phases to control surface tension, drying, or viscosity. Such inks were used to print 3D objects from composite materials or even integrated Li-ion batteries~\cite{Kokkinis2015,Sun2013}.

As described above, non-aqueous solvents are used in many practical applications. However, there is only scarce data in the literature on interactions between solid surfaces across non-aqueous polar media and their mixtures with water. Forces between mica sheets with SFA across polar propylene carbonate, acetone, methanol, and ethylene glycol were first measured by Christenson and Horn~\cite{Christenson1983,Christenson1984,Christenson1985}. In these measurements, two regimes were observed, the long-range behavior, which was dominated by repulsive double-layer force, and the short-range behavior, which included oscillatory forces. These oscillations are formed by structuring of solvent molecules near the solid surface~\cite{Christenson1983, Smith2017a}. The forces between silica surfaces in ethylene glycol were measured with colloidal probe technique~\cite{Atkins1997a}. At large distances, long-range repulsion was observed, and at short distances, hydration-like repulsion was measured. Attractive solvation interactions were measured when fluorocarbon surfaces were interacting across ethylene glycol~\cite{Parker1992a}. The above cited research has shown that double-layer and solvation forces similar to ones measured in water are also present in non-aqueous polar media.

Forces between silica surfaces across alcohols and alcohol-water mixtures were investigated with a colloidal probe technique~\cite{Kanda1998,Kanda1999,McNamee2001,Franz2002,Govrin2017}. In these systems, the range and magnitude of the double-layer force change by changing water content in the mixture. The variation of the decay length of double-layer forces was attributed to the variation of the dielectric constant in the mixtures and ion association at high alcohol contents~\cite{Kanda1998}. At short distances a step-like repulsion was observed in pure alcohols~\cite{Kanda1998,Kanda1999,Franz2002}. These short-range forces stem form the ordering of the alcohol molecules near the surface.

Lately, there has been a lot of interest in surface forces across ionic liquids and highly concentrated aqueous salt solutions. It has been shown that in very concentrated and pure ionic liquids long-range exponential repulsion exists~\cite{Gebbie2013,Smith2016,Smith2017a}. These repulsions have decay lengths much larger compared to the Debye length and furthermore the decay length increases with increasing concentration in highly concentrated systems. This behavior is opposite to the behavior observed in dilute electrolytes~\cite{Smith2016}. It was further observed that at high concentrations of electrolytes the transition from monotonic to oscillatory forces is present~\cite{Smith2016} and that the wavelength of these oscillations can be abruptly changed by varying solvent composition~\cite{Smith2017a}. These experiments sparked a renewed interest in theoretical description of ionic fluids and a variety of theoretical approaches were utilized to describe these new exciting experimental data~\cite{Coupette2018,Lee2017,Kjellander2018,Kjellander2019,Adar2019,Avni2020,Coles2020}. Measurements mentioned above have been performed using SFA, where one typically uses mica as a surface. Using colloidal probe AFM would enable to use other surfaces during similar experiments, which would test the hypothesis that the phenomena observed in ionic liquids are interface independent.

Here, we investigate forces between silica colloids in isopropanol solutions of 1:1 electrolytes. Due to a lower dielectric constant of isopropanol as compared to water, the electrostatic coupling is stronger in these solutions. The stronger electrostatic coupling induces a very rich behavior of these simple systems. Variety of different type of forces are found. In addition to double-layer and van der Waals interactions, attractive non-DLVO and short-range solvation forces are present. At increased concentrations the transition from monotonic exponential to damped oscillatory interactions is observed.

\section{Materials and Methods}

\subsection{Force Measurements}

\begin{figure}[t]
\centering
\includegraphics[width=4.75cm]{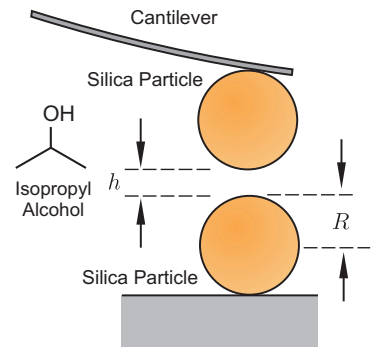}
\caption{Schematic representation of a colloidal probe experiment. Force measurements between two silica particles were done in isopropanol solutions.}
\label{fig:colloidal_probe}
\end{figure}

The surface force measurements were performed on a closed-loop atomic force microscope (MFP-3D, Asylum Research) using the colloidal probe technique at room temperature $23\pm2$~\celsius, see Fig.~\ref{fig:colloidal_probe}. The AFM is mounted on an inverted optical microscope (Olympus IX70). Spherical silica 4~$\mu m$ particles (Bangs Laboratories Inc., USA) were attached to tipless cantilevers (MikroMasch, Tallin, Estonia) with the help of a small amount of glue (Araldite 2000+). Some particles were spread onto quartz polished disk (Robson Scientific, Sawbridgeworth, UK) which was used as a bottom of a liquid cell in which measurements were done. The quartz disk was beforehand cleaned in piranha solution (3:1 mixture of H$_2$SO$_4$ 98\% and H$_2$O$_2$ 30\%). Both cantilevers with particles and the substrate were heated in an oven at 1200~\celsius\ for 2 hours, to burn the glue, achieve firm attachment and decrease surface roughness of the particles. Solutions were made in isopropanol (99.8\%, Extra Dry, AcroSeal, Acros Organics) with addition of tetrabutylammonium bromide (TBAB, 99+\%, Acros Organics) or lithium chloride (LiCl, BioXtra, $>$99.0\%, Sigma-Aldrich).
%%%%%%%%%%%%%%%%%%%%%%%%%%%%%%%%%%%%%
%%%%%%%%%%%%%%%%%%%%%%%%%%%%%%%%%%%%%

Before force measurement, cantilevers and substrate were cleaned in Milli-Q water and ethanol, and then treated in plasma for 20 minutes. When mounted into the AFM, the fluid cell was filled with solution, and the particle on the cantilever was centered above another one on the quartz disk with a precision of around 100~nm. The speed of the approach of the cantilever to the substrate during the measurement was 500~nm/s for all experiments, except for 50~mM solutions, where the approach speed of 100~nm/s was used. For a selected pair, the cantilever deflection was recorded in 150 approach-retract cycles. For each salt concentration, measurement was done on 3-5 different pairs of particles. The approach parts of the curves were averaged and used for analysis. Hooke’s law was used to convert deflection to force. Cantilever spring constant was determined by the Sader method~\cite{Sader1999}.

\subsection{Analysis of the Force Curves}

The extended DLVO theory is used to analyze the force curves~\cite{Israelachvili2011, Russel1989}
\begin{equation}
F = F_{\rm vdW} + F_{\rm dl} + F_{\rm att} ,
\label{eq:dlvo}
\end{equation}
where the total force between two particles is a superposition of van der Waals, $F_{\rm vdW}$, and double-layer, $F_{\rm dl}$, forces as in classical DLVO and we add an additional attractive exponential term, $F_{\rm att}$. The van der Waals force is calculated with a non-retarded expression for two spherical particles with radius $R$~\cite{Israelachvili2011, Russel1989}
\begin{equation}
F_{\rm vdW} = -\frac{HR}{12}\cdot \frac{1}{h^2} ,
\label{eq:vdw}
\end{equation}
where $H$ is the Hamaker constant and $h$ is the surface-surface separation.

The double-layer forces are calculated by solving the Poisson-Boltzmann equation in the plate-plate geometry
\begin{equation}
\frac{{\rm d}^2 \psi(x)}{{\rm d} x^2} =  \frac{2e_0c}{\varepsilon\varepsilon_0}\sinh (\beta e_0 \psi) ,
\label{eq:pb}
\end{equation}
where $e_0$ is the elementary charge, $c$ is the number concentration of the 1:1 electrolyte, $\beta = 1/(\kT)$ is the inverse thermal energy, $\varepsilon_0$ is the vacuum permittivity, and $\varepsilon = 17.9$ is the dielectric constant of the isopropanol. $\psi(x)$ is the electric potential, and $x$ is the coordinate normal to the plates. The plates are positioned at $x = -h/2$ and $x = h/2$. Due to symmetry, the Poisson-Boltzmann equation is solved in the $0 \le x \le h/2$ half-space with the following boundary conditions
\begin{gather}
\left. \frac{\mathrm{d} \psi }{\mathrm{d} x} \right|_{x=0} = 0 \quad {\rm and} \\
\left. \varepsilon\varepsilon_0\frac{\mathrm{d} \psi }{\mathrm{d} x} \right|_{x=h/2} = \sigma - C_{\rm in}[\psi(h/2) - \psi_{\rm dl}] 
\, ,
\end{gather}
where $\sigma$ and $\psi_{\rm dl}$ are surface charge density and diffuse-layer potential of the isolated surface, respectively. These two parameters are connected through
\begin{equation}
\sigma = \frac{2\kappa \varepsilon\varepsilon_0}{\beta e_0} \sinh\left( \frac{\beta e_0 \psi_{\rm dl}}{2}\right) ,
\label{eq:charge}
\end{equation}
where $\kappa$ is the inverse Debye length
\begin{equation}
\kappa=\sqrt{\frac{2\beta e_0^2 c}{\varepsilon\varepsilon_0}} .
\label{eq:debye-length}
\end{equation}
$C_{\rm in}$ is the inner-layer capacitance. The regulation parameter, $p$, is used for the interpretation of capacitances. This parameter interpolates between constant potential (CP) with $p=0$ and constant charge (CC) with $p=1$, and it is defined as
\begin{equation}
p = \frac{C_{\rm dl}}{C_{\rm dl} + C_{\rm in}} ,
\label{eq:regulation}
\end{equation}
where diffuse-layer capacitance, $C_{\rm dl}$, is calculated as
\begin{equation}
C_{\rm dl} = \varepsilon\varepsilon_0 \kappa \cosh \left( \frac{\beta e_0 \psi_{\rm dl}}{2}\right) .
\label{eq:dl-capacitance}
\end{equation}
The solution of the Poisson-Boltzmann \eq~(\ref{eq:pb}) yields the electric potential profile between two surfaces, $\psi(x)$, from which its value at the mid-plane can be extracted $\psi_{\rm M} = \psi(0)$. The disjoining pressure is then calculated as
\begin{equation}
\Pi (h) = 2\kT c \left[ \cosh(\beta e_0 \psi_{\rm M}) - 1 \right] .
\label{eq:pressure}
\end{equation} 
The pressure is then integrated to obtain energy per unit area for two plates
\begin{equation}
W_{\rm dl} = \int_h^{\infty} \Pi (h') {\rm  d} h'  .
\label{eq:energy}
\end{equation}
The double-layer force between two spherical particles of radius, $R$, is then obtained by using the Derjaguin approximation, which connects sphere-sphere and plate-plate geometries~\cite{Israelachvili2011, Russel1989} 
\begin{equation}
F_{\rm dl} = 2\pi R_{\rm eff} W_{\rm dl} ,
\label{eq:derjaguin}
\end{equation}
where $R_{\rm eff}$ is the effective radius and is equal to $R/2$ for two identical spheres. 

The non-DLVO additional attractive term defined in \eq~(\ref{eq:dlvo}) is modeled with an exponential function~\cite{Moazzami-Gudarzi2016c, Kanduc2017}
\begin{equation}
F_{\rm att} = -AR_{\rm eff}e^{-qh} ,
\label{eq:attraction}
\end{equation}
where $A$ is the amplitude and $q^{-1}$ is the decay length of this additional force.

At higher concentrations of salt, damped oscillatory forces are present and they are modeled as
\begin{equation}
F_{\rm osc} = BR_{\rm eff}e^{-h/\xi}\cos (2\pi/\lambda + \phi) ,
\label{eq:oscillations}
\end{equation}
where $B$ is the amplitude, $\xi$ is the decay of damping, $\lambda$ is the wavelength, and $\phi$ is the phase shift of the oscillations.

\section{Results and Discussions}

Colloidal probe technique based on AFM is used to measure the forces between silica colloids across alcohol solutions. Specifically, we study the interactions in tetrabutylamonium bromide (TBAB) and LiCl solutions. Both salts are 1:1 electrolyte dissolved in isopropanol.

\begin{figure}[t]
\centering
\includegraphics[width=8.5cm]{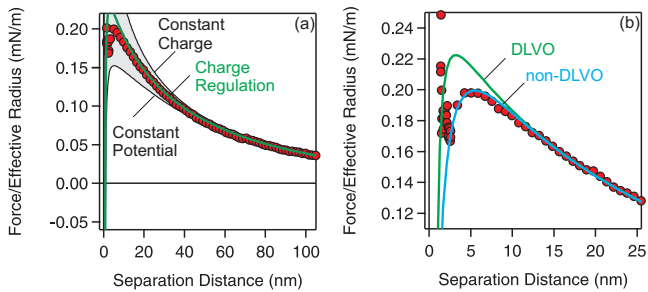}
\caption{Forces in isopropanol without added electrolyte. (a) DLVO fits with constant charge (CC), constant potential (CP), and constant regulation (CR). (b) Comparison of DLVO and non-DLVO fits with additional exponential attraction. Note that in both cases the Hamaker constant $H = 1.0\cdot 10^{-21}$~J is fixed in the fits.}
\label{fig:forces_pure}
\end{figure}

First, we look at the interactions between silica particles in isopropanol without added salt, which are presented in Fig.~\ref{fig:forces_pure}.
%%%%%%%%%%%%%%%%%%%%%%%%%%%%%%%%%%%%%
%%%%%%%%%%%%%%%%%%%%%%%%%%%%%%%%%%%%%
Without added salt forces are repulsive and long-ranged with a decay length of $\sim 80$~nm. They can be accurately fitted with DLVO theory with constant regulation approximation down to separation distance of about 10~nm, see Fig.~\ref{fig:forces_pure}a. This fit allows to extract the diffuse-layer potential, the regulation parameter, and the electrolyte concentration. The extracted diffuse-layer potential is equal to 101~mV and can be converted through \eq~(\ref{eq:charge}) to diffuse-layer surface charge density of 0.34~mC/m$^2$. The latter value is about 10-20 times lower as compared to silica in water~\cite{Stojimirovic2020,Uzelac2017,Valmacco2016a}. The lower charge density of the silica surface can be explained by longer-range electrostatics in solvents with lower dielectric constants. Bjerrum length, which estimates the distance at which electrostatic interaction is equal to the thermal energy, is equal to 0.71~nm in water, while it equals to 3.1~nm in isopropanol. These values show, that more energy is needed to separate a negative and a positive charge in alcohol and therefore it is harder to charge surface in alcohol solutions. The regulation parameter determined from the force, $p = 0.52$, suggests that the charge regulation of silica surfaces upon approach is considerable. Further, we assume 1:1 electrolyte to be present in the alcohol solution, where the fitted concentration is equal to 3.2~$\mu$M. These traces of ions present in the solutions are possibly coming from the small amount of water in alcohol. Note that we did not tried to remove traces of ions before the measurements.

A more detailed graph of interaction in pure isopropanol, shown in Fig.~\ref{fig:forces_pure}b, reveals that DLVO theory overestimates the force at distances lower that $\sim 10$~nm. Therefore below 10~nm attractive non-DLVO forces are present. This additional attraction can be modeled with simple exponential attraction described in \eq~(\ref{eq:attraction}). The improved non-DLVO model is accurate down to separations of about 1~nm. The extracted decay length and amplitude of the additional attraction are equal to $q^{-1} = 2.2$~nm and $A=0.16$~mN/m, respectively. These additional non-DLVO forces will be addressed in more detail below.

\begin{figure}[t]
\centering
\includegraphics[width=8.5cm]{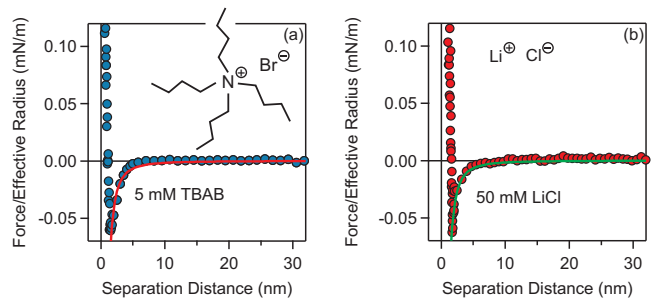}
\caption{Van der Waals forces between silica in isopropanol solutions with (a) 5 mM tetrabutylamonium bromide (TBAB) and (b) 50 mM of lithium chloride (LiCl).}
\label{fig:vdw}
\end{figure}

Let us now look at forces at high concentrations of added salt. We refer here to high salt concentration, for conditions when the double-layer interactions are completely screened and van der Waals attraction is dominant. In isopropanol these conditions are reached already at 5~mM of added salt, which is about a factor of 10-100 lower as compared to aqueous systems~\cite{Stojimirovic2020, Valmacco2016a, Uzelac2017, Dishon2009}. In Fig.~\ref{fig:vdw} the van der Waals interaction between silica is shown for the two salts.
%%%%%%%%%%%%%%%%%%%%%%%%%%%%%%%%%%%%%
%%%%%%%%%%%%%%%%%%%%%%%%%%%%%%%%%%%%%
In both situations, the attraction can be accurately fitted with \eq~(\ref{eq:vdw}) and the Hamaker constant of $H = (1.0\pm 0.1)\cdot 10^{-21}$~J can be extracted. This constant is slightly lower than the one measured across aqueous solutions for similar particles~\cite{Uzelac2017}. This difference is due to higher refractive index of isopropanol as compared to water. The low value of the Hamaker constant is probably also a consequence of some residual nanoscale roughness of the particles. The Hamaker constant has been shown to decrease with increasing roughness.~\cite{Thormann2017, Valmacco2016}. The van der Waals interaction does not depend on the type of added salt, which is consistent with earlier observations~\cite{Moazzami-Gudarzi2016c, Smith2020}.

\begin{figure}[t]
\centering
\includegraphics[width=8.5cm]{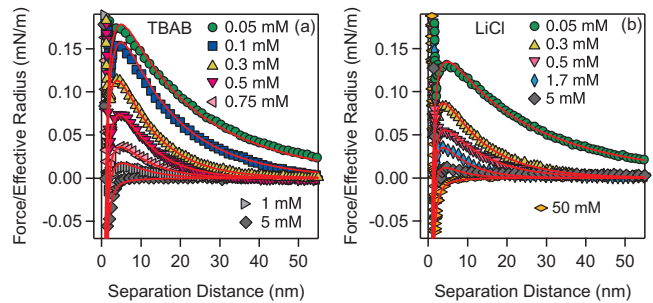}
\caption{Forces at different concentrations of (a) tetrabutylamonium bromide (TBAB) and (b) lithium chloride (LiCl).}
\label{fig:forces_concentration}
\end{figure}

The forces for the transition from low to high salt are shown in Fig.~\ref{fig:forces_concentration}.
%%%%%%%%%%%%%%%%%%%%%%%%%%%%%%%%%%%%%
%%%%%%%%%%%%%%%%%%%%%%%%%%%%%%%%%%%%%
The experiments were performed in two different salt solutions, namely tetrabutylamonium bromide (TBAB) and lithium chloride (LiCl). These measurements enable us to study the influence of ion size on double-layer interactions, since the tetrabutylamonium (TBA$^+$) ion is bulkier compared to the lithium ion. In both cases the transition from repulsive to attractive forces is observed by increasing salt concentration. The repulsive forces are slightly longer-ranged in the case of LiCl as compared to TBAB. Furthermore, a higher concentration of LiCl is needed to completely screen the repulsion as compared to TBAB. In order to extract more details about these systems, we have fitted the experimental curves with extended DLVO theory, see \eq~(\ref{eq:dlvo}). In these fits the Hamaker constant was fixed to the value of $H = 1.0\cdot 10^{-21}$~J, which is consistent with high salt measurements. Diffuse-layer potential, electrolyte concentration, regulation parameter, and additional attraction amplitude and decay were determined by least square fitting. The extended DLVO theory accurately describes the experimental curves down to the separation distances of $\sim 1$~nm.

Regulation parameters were observed to be independent of concentration and equal to $0.51\pm 0.08$ and $0.62\pm 0.14$ for TBAB and LiCl, respectively. For both salts regulation parameters are similar to the values obtained in pure isopropanol and the surfaces regulate fairly strongly.

\begin{figure}[t]
\centering
\includegraphics[width=8.5cm]{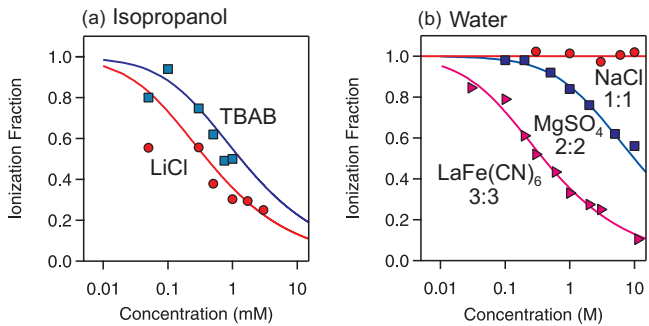}
\caption{(a) Ionization fractions of TBAB and LiCl salts in isopropanol as determined from AFM force measurements. (b) Ionization fractions of 1:1, 2:2, and 3:3 salts in water, data taken from~\cite{Smith2019}. The solid lines are calculated with chemical equilibrium model shown in \eq~(\ref{eq:association_constant}).}
\label{fig:ionization_fraction}
\end{figure}

The concentration of free ions can also be determined from the double-layer fitting. All the extracted concentrations were smaller than nominal ones for both salts investigated. Therefore the measured decay lengths of the double-layer forces are larger than expected based on nominal salt concentrations. These results suggest that the salts are not fully dissociated and that some fraction of ions form ion pairs~\cite{Smith2019, dosSantos2010a}. The ionization fractions for both salts in isopropanol are plotted as a function of nominal salt concentrations in Fig.~\ref{fig:ionization_fraction}a. 
%%%%%%%%%%%%%%%%%%%%%%%%%%%%%%%%%%%%%
%%%%%%%%%%%%%%%%%%%%%%%%%%%%%%%%%%%%%
The ionization fractions are approaching unity only for very dilute isopropanol solutions and are rapidly dropping at concentrations above 0.1~mM. Above 1~mM more than 50~\% of ions form ion pairs. This behavior can be very well explained by the chemical equilibrium model, which accounts for ion pair formation
\begin{equation}
{\rm A^+ + B^- \rightleftharpoons AB} ,
\label{eq:equilibrium}
\end{equation}
where A$^+$ and B$^-$ are cations and anions, respectively, while AB represents a neutral ion pair. This equilibrium can be quantified by the following mass action law
\begin{equation}
K = \frac{[{\rm A^+}][{\rm B^-}]}{\rm [AB]} ,
\label{eq:association_constant}
\end{equation}
where $K$ is the association constant and square brackets denote molar concentrations. The concentration of free ions is equal to $c_{\rm free} = {\rm [A^+] = [B^-]}$ and total concentration is $c_{\rm tot} = {\rm [A^+] + [AB]}$. The ionization fraction is finally defined as a ratio $c_{\rm free}/c_{\rm tot}$. The solid lines in Fig.~\ref{fig:ionization_fraction}a are calculated with the chemical equilibrium model \eq~(\ref{eq:association_constant}), where the equilibrium constant $K$ is the only adjustable parameter. The fitted association constants for TBAB and LiCl are 1.5~mol/L and 5.0~mol/L, respectively. In literature these values are typically represented as logarithmic form $\log_{10} K$, which gives values of 3.2 and 3.7 for TBAB and LiCl, respectively. The association constant can be independently determined from electrical conductivity measurements by analysis developed by Fuoss and Onsager~\cite{Fuoss1957,Fuoss1958}. For TBAB association constants were determined in methanol and ethanol mixtures~\cite{Sadek1959}. The $\log_{10} K$ values for the dielectric constant corresponding to the present isopropanol system ($\varepsilon = 17.9$) are 2.4 and 2.8 for methanol and ethanol based solutions, respectively~\cite{Sadek1959}. Our value for TBAB in isopropanol of $\log_{10} K = 3.2$ is therefore perfectly consistent with the published results on TBAB methanol and ethanol solutions.

Ion pairing cannot be completely understood by accounting only for electrostatic interactions and additional solvent specific interactions must be taken into account~\cite{Sadek1959}. However, Bjerrum theory which includes only Coulombic and hard-sphere interactions still gives reasonable estimates for the the extent of ion association. According to Bjerrum theory, the association constant can be calculated as~\cite{Valeriani2010, Bjerrum1926}
\begin{equation}
K = 4\pi N_{\rm A} \int_{r_{\rm min}}^{r_{\rm max}} e^{-\beta U(r)}r^2 {\rm d}r ,
\label{eq:bjerrum}
\end{equation}
where $N_{\rm A}$ is the Avogadro number, $r$ is the center to center distance of the ions, and $U(r) = -\ell_{\rm B}/r$ is the electrostatic energy between cation and anion with $\ell_{\rm B} = e_0^2/(4\pi \varepsilon\varepsilon_0)$ being the Bjerrum length. The bounds of the integral are the minimal distance the two ions can approach, $r_{\rm min}$, and the maximal distance at which we consider ions to be paired, $r_{\rm max}$. While the minimal approach distance is determined by ion size, the maximal distance is less defined, however the precise value of the upper bound does not affect the results drastically~\cite{Valeriani2010}. The Bjerrum theory permits us to estimate the ion sizes based on the constants extracted from ionization fractions. The calculated values of minimal approach are 2.5~\AA\ for LiCl and 3.0~\AA\ for TBAB. The former value agrees perfectly with the sum of the Li$^+$ (0.7~\AA) and Cl$^-$ (1.8~\AA) radii~\cite{Marcus1988a}. While the reported values of minimal approach distance for tetrabutylamonium salts in non-aqueous solvents vary substantially~\cite{Valeriani2010}, the average of $\sim 3$~\AA\ agrees well with our result for TBAB. The difference in the ionization fraction for LiCl and TBAB in alcohol is therefore due to the difference in ion size. The bulkier TBAB salt forms less ion pairs compared to more compact LiCl.

One can further compare ion association in isopropanol with pairing in aqueous systems, see Fig.~\ref{fig:ionization_fraction}b. In water, the 1:1 electrolytes do not show any ion pairing, and association only becomes prominent in the 2:2 and 3:3 electrolytes~\cite{Smith2019, Marcus2006}. The ionization fractions measured for 1:1 electrolyte in isopropanol is somewhere between the values measured for 2:2 and 3:3 electrolytes in water. This fact can be understood by comparing Bjerrum lengths in water (0.71~nm) and in isopropanol (3.1~nm). Since the Bjerrum length in water is about 4-5 times smaller than in isopropanol, the ions have to be more charged to achieve the same electrostatic attraction energy at contact.

\begin{figure}[t]
\centering
\includegraphics[width=8.5cm]{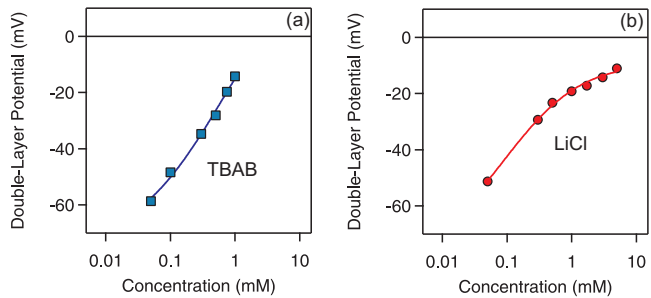}
\caption{Surface potentials extracted from AFM force measurements and electrokinetic measurements in alcohol solutions of (a) TBAB and (b) LiCl.}
\label{fig:surface_potential}
\end{figure}

The diffuse-layer potentials extracted from the force curves are show in Fig.~\ref{fig:surface_potential}.
%%%%%%%%%%%%%%%%%%%%%%%%%%%%%%%%%%%%%
%%%%%%%%%%%%%%%%%%%%%%%%%%%%%%%%%%%%%
In both solutions the potentials increase with increasing concentration as they are progressively screened by adding more ions in the solution. The values for both salts are similar at low concentrations, while at higher concentrations the screening of TBAB is more effective. This difference stems from the more effective dissociation of TBAB in solutions. One can convert the potentials to surface charge by the means of \eq~(\ref{eq:charge}). Note that in this equation a concentration of free ions and not nominal concentration of salt has to be used. The resulting average surface charge densities for TBAB and LiCl solutions are $-0.50\pm 0.10$ and $-0.40\pm 0.04$ mC/m$^2$, respectively. The slightly lower magnitude of the surface charge density for LiCl solutions is possibly connected to the stronger association of the Li$^+$ ion with the negatively charged silanol groups. This association is probably less prominent for TBA$^+$, and therefore, this ion is less effective in neutralizing the surface charge.

\begin{figure}[t]
\centering
\includegraphics[width=8.5cm]{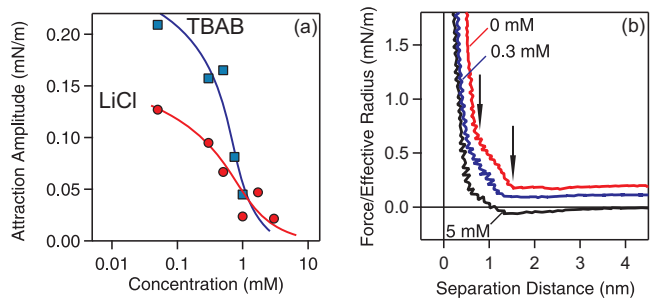}
\caption{(a) Amplitude of the additional non-DLVO exponential attraction in isopropanol solutions. (b) Short-ranged repulsive non-DLVO forces in TBAB solutions, due to isopropanol structuring near the surface. Arrows indicate the steps in the force profile.}
\label{fig:non_dlvo}
\end{figure}

Finally, let us look at the non-DLVO interactions observed in alcohol solutions. Similarly, the non-DLVO attractions in isopropanol without added salt, shown in Fig.~\ref{fig:forces_pure}b, such attractions are also present in both TBAB and LiCl solutions. These attractions can be described by the decay length, $q^{-1}$, of 2.2~nm and 2.5~nm for TBAB and LiCl, respectively. The fitted amplitudes of the attractions, $A$, defined in \eq~(\ref{eq:attraction}), are shown in Fig.~\ref{fig:non_dlvo}a.
%%%%%%%%%%%%%%%%%%%%%%%%%%%%%%%%%%%%%
%%%%%%%%%%%%%%%%%%%%%%%%%%%%%%%%%%%%%
The attractions are the strongest at low salt levels and they disappear at concentrations above $\sim 1$~mM. Non-DLVO attractions between silica surfaces are not present in the aqueous solutions of simple monovalent electrolytes, like KCl~\cite{Uzelac2017,Valmacco2016a, Smith2018a}. On the other hand they were observed in aqueous solutions of hydrophobic monovalent ions~\cite{Smith2018}, and in the presence of multivalent counterions~\cite{Moazzami-Gudarzi2016c, Valmacco2016a, Kanduc2017}. In aqueous solutions of monovalent hydrophobic ions the decay lengths of these attractions are between 1.5 and 3~nm, while they are around 1~nm in the solutions of multivalent counterions. Similar to the present case of alcohol solutions, these attractive non-DLVO forces also disappear in water at high salt concentrations. Currently the source of these attractive forces in aqueous media is not clear as they might be connected to ion-ion correlation~\cite{Kanduc2017}, lateral charge heterogeneities~\cite{Smith2020}, spontaneous charge fluctuations~\cite{Adzic2015}, or possibly varying dielectric constant close to the surface. However, these non-DLVO attractions seem to be present, when ions strongly interact with the surface~\cite{Smith2020}. The presence of these forces in alcohol solutions seems to confirm this observation, since due to lower dielectric constant the electrostatic interaction between the ions and the surface is enhanced. In water, the interaction between ions and surface is strong enough only in the case of multivalent counterions, or if ions interact through other strong non-electrostatic interactions, for example hydrophobic force.

%%%%%%%%%%%%%%%%%%%%%%%%%%%%%%%%%%%%%
\begin{figure}[t]
\centering
\includegraphics[width=8.5cm]{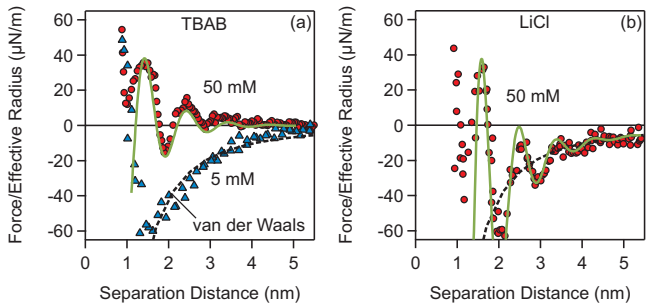}
\caption{Oscillatory forces observed at 50~mM electrolytes (a) TBAB and (b) LiCl. The full lines represent the fit to \eq~(\ref{eq:oscillations}), while dashed lines show van der Waals interaction. Note that in the case of LiCl van der Waals force is added to the damped oscillatory force.}
\label{fig:oscillations}
\end{figure}
%%%%%%%%%%%%%%%%%%%%%%%%%%%%%%%%%%%%%

In addition to non-DLVO attraction, a short-range non-DLVO repulsion is also observed in alcohols, see Fig.~\ref{fig:non_dlvo}b. These forces have been observed in alcohols before~\cite{Franz2002, Kanda1998} and are caused by structuring of alcohol molecules close to the solid surface. When sharp AFM tip is used for the measurements the resulting profile is oscillatory with a period of about 0.95~nm in 1-propanol~\cite{Franz2002}. In our case, the oscillations are probably smeared out due to surface roughness of the colloidal probe, however the steps with the period of about 1~nm can be clearly observed in the sample without added salt. Upon addition of salt the structuring of alcohol molecules close to the surface seem to be disturbed and the steps become less clear. This disturbance of the alcohol layering is probably caused by adsorption of ions to the surfaces.

At increased salt concentration another type of interactions becomes evident. One can observe a transition between monotonic interaction and damped oscillatory force. In Fig.~\ref{fig:oscillations} this transition is shown for TBAB salt. At 5~mM, monotonic van der Waals interaction is present, while oscillations in the force profile become clearly evident at 50~mM. Similarly, an oscillatory profile on top of van der Waals attraction is observed in 50~mM LiCl shown in Fig.~\ref{fig:oscillations}b. We suspect that the transition from monotonic to the oscillatory behavior is the Kirkwood cross-over~\cite{Carvalho1994,Kirkwood1939a}. Recently, different theoretical and simulation approaches were used to study this transition~\cite{Carvalho1994, Kjellander2019, Adar2019, Coupette2018,Avni2020,Coles2020, Keblinski2000}. These approaches predict the transitions at salt concentrations corresponding to $\kappa d~\sim 1 - 2$, where $d$ is the mean ion diameter. In the present case we observe the transition below $\kappa d \sim 0.5$. This shift to lower concentrations, might be connected to the strong electrostatic coupling, which is present in the current alcohol system. However, dressed-ion theory predicts the shift to larger $\kappa d$ values for the 2:2 electrolytes in water, where the electrostatic coupling is also stronger~\cite{Ennis1995}. The origin of the observed shift of the monotonic to oscillatory transition in alcohol solutions is therefore not clear and could be also caused by solvent molecules~\cite{Coupette2018}.

The Kirkwood cross-over was also observed in aqueous solutions of simple ions and solutions of ionic liquids, albeit one has to increase the concentrations beyond few molar in these systems~\cite{Smith2016,Smith2017a}. In the present alcohol case this cross-over occurs at concentrations of few tens of mM, which is at two orders of magnitude lower concentrations. These low concentrations provide a larger window for exploration of these effects in future.

In order to extract the wavelength of the oscillations, $\lambda$ and decay of exponential damping $\xi$, we model these forces with \eq~(\ref{eq:oscillations}). The fitted wavelengths, $\lambda$, for TBAB and LiCl are equal to $1.0\pm0.1$~nm and $0.9\pm0.1$, respectively. The decay of the oscillations, $\xi=0.65\pm0.1$, is the same for both salts. The observed wavelength is slightly lower for LiCl as compared to TBAB, but in both cases the wavelength is about two times larger than the diameter of the ion pair. The corresponding wave length in 2~M NaCl aqueous solution was reported to be $\sim 0.5$~nm, which is about the size of an ion pair~\cite{Smith2016}. Similarly, the wavelength in ionic liquid-solvent mixtures was measured to be about the ion pair size~\cite{Smith2017a}. While currently we have no explanation, why in the present alcohol solutions, the wavelength is about two times larger than diameter of the ion pair, this observation might be again connected with strong electrostatic coupling in alcohol solutions. Further theoretical studies would be needed to understand Kirkwood cross-over and oscillatory forces in these strongly electrostatically coupled systems.

\section{Conclusions}

Forces between negatively charged silica particles were measured in monovalent salt solutions in isopropanol. An extremely rich behavior of these systems is observed; this includes, van der Waals forces, double-layer forces, attractive non-DLVO forces, repulsive solvation forces, and damped oscillatory interactions at increased concentrations. The richness of these systems is connected to strong electrostatic coupling, which is due to low dielectric constant of isopropanol.

The interactions between silica surfaces are repulsive at low salt levels and become progressively more attractive with increasing salt concentration. This behavior is consistent with DLVO theory. However, the decay of the double-layer repulsion is much longer than expected from Debye lengths calculated from nominal salt concentrations. This observation can be explained by ion pairing and quantified by Bjerrum theory. The association constants for the 1:1 electrolyte in alcohol are comparable to the association constants in the 2:2 and 3:3 aqueous electrolytes, since Bjerrum lenghts in isopropanol are about 4-5 times larger as compared to water solutions.

At distances below $\sim 10$~nm the experimental force profiles deviate from DLVO theory, and additional non-DLVO forces are observed. The additional attractive forces are possibly caused by surface charge heterogeneities, which are induced by strong ion adsorption. This strong adsorption is driven by strong electrostatic interaction between ions in solution and charged surface groups.

At distances below 2~nm repulsion due to structuring of isopropanol molecules close to the surface is present. This structuring is disturbed with increasing salt concentration, due to adsorption of counterions.

Finally, the transition from monotonic to oscillatory forces is observed at increased concentrations. We believe that this observation is the consequence of the Kirkwood cross-over, albeit the concentration of this cross-over seems to be lower than predicted by theoretical studies and found in aqueous systems. Furthermore, the wavelength of the oscillations is longer than what is expected from the diameter of the ion pair. These differences might be connected to strong electrostatic coupling in alcohol systems. We hope that the present experimental data will enable testing of recent theoretical approaches and spark the interest for more detailed exploration of phenomena found in these systems.

\section*{Acknowledgments}
This research was supported by the Swiss National Science Foundation through grant 162420 and the University of Geneva. The authors are thankful to Michal Borkovec for insightful discussions and for providing access to the instruments in his laboratory and to Plinio Maroni for the help with AFM measurements and Poisson-Boltzmann solution software.

\bibliography{forcesAlcohols.bib}
\bibliographystyle{apsrev4-1}

\end{document}